# Computationally efficient optic nerve head detection in retinal fundus images

Reza Pourreza-Shahri[2], Meysam Tavakoli[1], Nasser Kehtarnavaz[2]

Oklahoma State University, University of Texas at Dallas

*Abstract*—This paper presents a computationally efficient method for the detection of optic nerve head in both color fundus and fluorescein angiography images. It involves a combination of Radon transformation and multi-overlapping windows within an optimization framework in order to achieve a robust detection in the presence of various structural, color, and intensity variations in such images. Three databases have been examined and it is shown that the introduced method provides high detection rates while achieving faster proceeding rates than the existing algorithms that possess comparable detection rates.

## I. INTRODUCTION

Computer Assisted Diagnosis (CAD) of retinopathy is currently being used to lower the workload of ophthalmologists as it is providing a non-labor intensive approach to the detection of the anatomical landmarks and lesions in retinal images. The computationally efficiency of such a CAD system would allow the screening of more patients within existing clinic or hospital time constraints.

The localization of retinal landmarks in particular Optic Nerve Head (ONH) constitutes the initial step towards identifying pathological conditions. Detection of ONH plays a key role in Diabetic Retinopathy (DR) [1, 2]. ONH is a yellowish region in a color fundus image occupying one seventh of the fundus image [3] (see Fig. 1). The main characteristic of ONH is its rapid intensity variation due to dark blood vessels that are in the vicinity of ONH. ONH has three characteristics that have been used for its localization: (1) it appears as a bright disk nearly 1600μm in diameter; (2) large blood vessels enter it from above and below; and (3) blood vessels diverge from it.

In many approaches, the ONH diameter is used as a length reference to measure objects in retina [1, 4]. In [3], the relatively constant distance between ONH and fovea was used to estimate the location of the latter. In [5], masking of ONH improved the classification of exudates regions and decreased the false positive rate. However, as noted in [6], detection of ONH is challenging due to the discontinuity of its boundary caused by large vessels as well as its considerable color or intensity variations as a result of structures such as exudates. The change in the color, shape, and depth of ONH provides a pathological sign, in particular a sign of glaucoma [7]. Glaucoma is the second most common cause of blindness in the world [8].

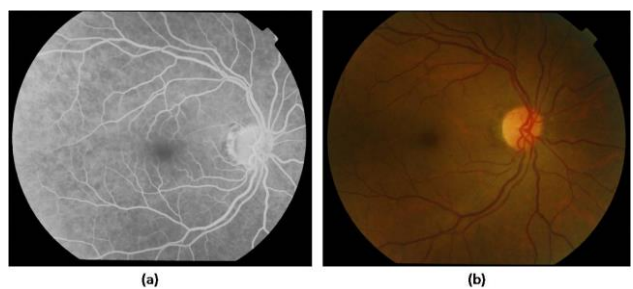

Fig. 1. Fundus image from same eye: (a) fluorescein angiography and (b) color fundus image.

## II. PREVIOUS WORKS

There are several existing algorithms that determine the location (generally center) of ONH or its boundary. Sinthanayothin *et al.* [3] used the area with the highest average intensity variation to detect ONH using an adaptive local contrast enhancement method. Walter and Klein [9] obtained the ONH center as the center of the brightest connected object in a fundus image instead of using intensity variation via the watershed transformation of the gradient image. Foracchia *et al.* [10] used the convergence of vessels to detect the ONH center. Youssif *et al.* [11] utilized the directional pattern of the retinal blood vessels for the detection of ONH. Their method involved normalizing luminosity and contrast throughout the image using illumination equalization and adaptive histogram equalization methods. Lu and Lim [12] located ONH based on its bright appearance in a color fundus image using a set of concentric lines with different directions and evaluated the image variation along multiple directions. The detection of ONH was achieved via the orientation of the line segment with the maximum or minimum variation.

A number of segmentation-based methods have also appeared in the literature. Li and Chutatape [13, 14] used an active shape model to detect ONH [15]. An active contour model was also discussed in [4, 16] by Osareh *et al.* [4, 16] to detect ONH. Lowell *et al.* [17] designed an ONH template and correlated it to the intensity component of the fundus image using the full Pearson-R correlation. Another model-based approach was presented by Xu *et al.*

in [18], where clustering-based classification of contour points was integrated into an active contour formulation. Wong *et al*. [19] used the level-set technique followed by ellipse matching. They obtained the ONH location by means of histogram analysis and a modified version of the conventional level-set method using the red channel.

Furthermore, Lu [20] designed a circular transformation to capture simultaneously both the circular shape of ONH and the image variation across the ONH boundary. A Hausdorff-based template matching together with a pyramidal decomposition were proposed by Lalonde *et al*. in [21]. Frank ter Haar [22] used illumination equalization in the green channel to address the difficulty of pyramidal decomposition in dealing with large areas of bright pixels.

Some ONH localization techniques not only use the ONH characteristics, but also exploit the location and orientation of vessels [23-27]. For example, Niemeijer *et al*. [27] presented the use of local vessel geometry and image intensity features. Tobin *et al*. [28] applied a method that mainly relied on vessels related to the ONH characteristics. A Bayesian classifier was used to classify each pixel in red-free images. Abramoff and Niemeijer [29] utilized the same ONH characteristics via kNN regression for the detection of ONH. The method introduced by Abramoff *et al*. [30] involved a pixel classification approach. In a recent study, Hsiao *et al*. localized ONH by an illumination correction operation, and contour segmentation via a supervised gradient vector flow snake [31]. Yu *et al* identified ONH candidates by first using template matching and then by using the vessel characteristics [32]. Finally, some techniques have made use of the fact that major retinal vessels converge into ONH [10, 11, 22, 33]. Many existing methods have reported noticeable failure for fundus images with a large number of white lesions, light artifacts or strongly visible choroidal vessels [34].

In this paper, a new ONH detection method based on Radon Transform (RT) and multi-overlapping windows is introduced. The key attribute of this new method is its computational efficiency. Two publicly available databases are examined: the DRIVE database [35], and the STARE dataset [36]. Noting that some retinal lesions, such as exudates, have intensity and color that are similar to ONH and thus can affect the performance of an automated ONH detection system, we have also considered DR images in color fundus images in the database MUMS-DB (Mashhad University of Medical Sciences image Database). In addition, fluorescein angiography (FA) fundus images in MUMS-DB are examined.

In the next section, the details of our algorithm for the detection of ONH are described.

### III. INTRODUCED METHOD

An RT-based algorithm is proposed here to detect ONH in fundus images. RT provides lower noise or intensity variation sensitivity due to the integration involved in it. A fundus image is first partitioned into overlapping blocks or sub-images. Local RT is then applied to each block or sub-image. The sub-images exhibiting peaks in the Radon space are then further processed in order to locate ONH.

The pipeline of the method is illustrated in Fig. 2.

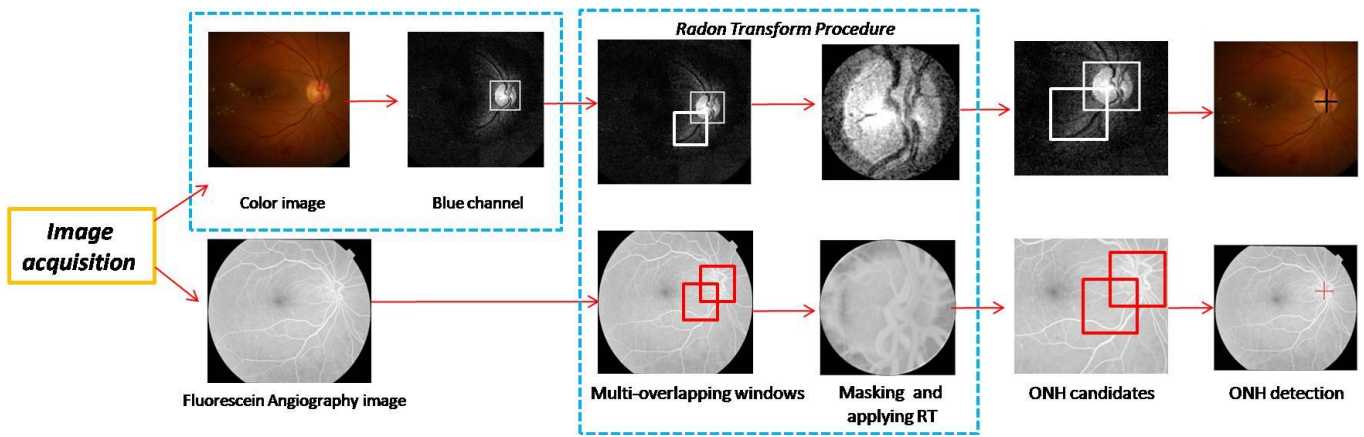

Fig. 2. Pipeline of the introduced method

The main components of the detection algorithm, the ONH detection procedure, and its computational efficiency aspect are covered in the following subsections A, B and C, respectively.

*A. Algorithm components*

The algorithm contains three main components as follows.

*1. Fundus region detection*

At first, it is important to separate fundus from background (or region that is out of the fundus field). A fundus image consists of a circular fundus and a dark background surrounding the fundus. Fundus is considered the region-of-interest and the processing is not applied to the background region. There are fundus masks that are supplied with retinal images in the databases examined. In a fundus mask, fundus pixels are marked with 1's and the background with 0's. With the help of the fundus mask, the detection algorithm would only process the pixels belonging to the fundus and not the background (see Fig. 3).

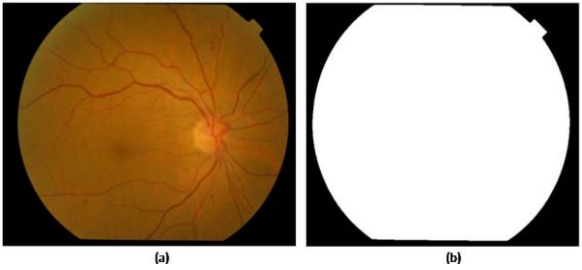

Fig. 3. (a) Fundus image, and (b) fundus mask

*2. Multi-overlapping window*

In our introduced method, a fundus image is partitioned into widows or sub-images. To find objects on the border of sub-images, overlapping sliding windows are considered. The size of the targeted object or ONH ($n$) is used to determine the size of the sub-image or sliding window. A parameter that has influence on the outcome is the windows overlapping ratio (*step*). If this ratio is equal to one, every pixel is examined just one time and sub-images would have no overlap. If the step is defined as two or more, then each pixel is examined ($n/step$) times either in horizontal or vertical sliding direction and this way each pixel is considered in up to $n^2$ sub-images (see Fig. 4).

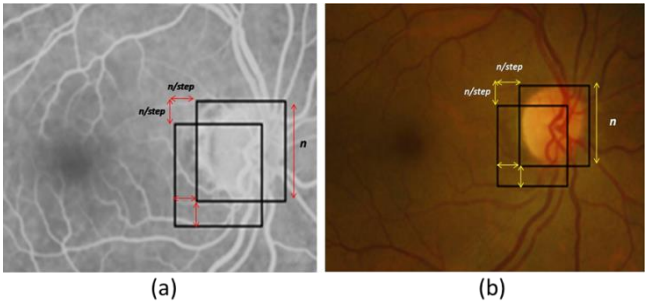

Fig. 4. Window size and overlapping ratio ($n$, $step$) in (a) fluorescein angiography, and (b) color fundus image

*3. Radon Transform*

Radon Transform (RT) is widely used in X-ray computed tomography (CT). A projection of a 2D function $f(x, y)$ corresponds to a set of line integrals. The Radon function computes the line integrals from multiple parallel paths or beams. The beams are spaced 1 pixel unit apart. To represent an image, the Radon function takes multiple, parallel-beam projections of the image from different angles by rotating the object of interest around the center of the image. Fig. 5 shows a single projection at a specified rotation angle.

Let us consider the RT of a planar function $f$:

$$\check{f}_\theta(s) = \int_{-\infty}^{+\infty} f(x,y)\,ds$$
$$= \int_{-\infty}^{+\infty} f\bigl(s(\cos\theta + \sin\theta) + z(-\sin\theta + \cos\theta)\bigr)\,ds \qquad (1)$$

where

$$\begin{bmatrix} x \\ y \end{bmatrix} = \begin{bmatrix} \cos\theta & \sin\theta \\ -\sin\theta & \cos\theta \end{bmatrix} \begin{bmatrix} s \\ z \end{bmatrix}$$

Then, the RT of a continuous two-dimensional function $f$ becomes

$$\check{f}_\theta(s) = \int_{-\infty}^{+\infty}\int_{-\infty}^{+\infty} f(x,y)\,\delta(s - x\cos\theta - y\sin\theta)\,dx\,dy \qquad (2)$$

A single projection of the object $\check{f}_\theta(s)$ is stated by Equation (2), where the Dirac delta function $\delta$ defines the path of the line integral. Equation (2) expresses the relationship between the object function $f(x, y)$ and the measured projection $\check{f}_\theta(s)$. The projection $\check{f}_\theta(s)$ in Equation (2) may be interpreted as the one-dimensional function $\check{f}_\theta(s)$ of a single variable $s$ with $\theta$ as a parameter. With the arrangement exhibited in Fig. 5, $\check{f}_\theta(s)$ is referred to as a parallel projection.

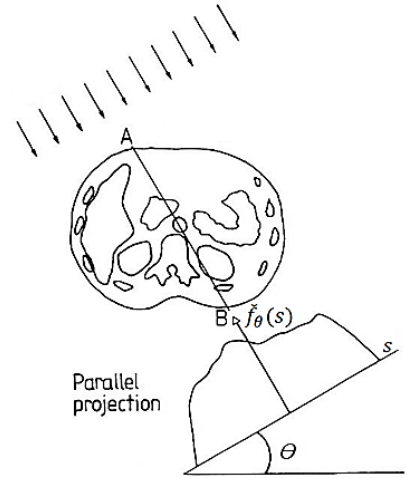

Fig. 5. Parallel projection line integral

RT is able to transform a pattern to a line in the Radon space and thus allows it to be easily distinguished from other patterns.

*B. ONH detection*

ONH is characterized by its bright circular characteristic and because of large vessels coming out and going into it, its edges are ill defined. ONH appears non-uniform in intensity, size, and location.

Our introduced algorithm for ONH detection in both FA and color fundus images comprises 4 steps:
1. Image partitioning
2. Applying Radon Transform
3. ONH validation
4. ONH detection

*Step 1) Image partitioning*

ONH is detected in local windows. The window size (*n*) has an effect on the detection outcome. Let

$$I = \{w(x_i, y_j) | i = 1 \dots N, j = 1 \dots M\} \qquad (3)$$

denote the sub-images $w(x_i, y_j)$ of size $n \times n$. If *n* is selected too small or too large, it would lead to extracting patterns other than ONH such as Microaneurysms or scar tissue. Based on our prior knowledge of the ONH size and its maximum diameter, an appropriate *n* is selected here. The size of *n* is chosen equal to the maximum diameter of ONH in a database. If the resolution of images is different, *n* is selected accordingly. For example, for the MUMS-DB database having an image resolution of 2896 × 1944, *n* is chosen to be 313 pixels, while for the DRIVE database, *n* is chosen to be 79 pixels and for the STARE database, *n* is chosen to be 130 pixels.

The window size is also determined automatically based on the scale factor of images under analysis. Given the actual size of ONH in mm, one can estimate the ONH size in pixels and subsequently the window size.

Another important parameter is the windows overlapping ratio (*step*). The processing speed depends on this *step* parameter. As *step* is increased, the computation time increases exponentially. In the results reported in this paper, *step* was chosen to be 4 meaning ¼ size window overlap.

*Step 2) Applying Radon Transform*

High intensity differences between ONH and background in FA and color images cause ONH to be associated with peaks in the Radon space. Moreover, in color images, the blue (B) component of sub-images is selected for applying RT to detect ONH due to its high contrast between its ONH and background. In the B channel, the red features like vessels, Microaneurysms, and Hemorrhages have low contrast and only yellowish pattern of ONH or exudates have high contrast. In other words, the B component provides a good separation for the white and yellow colors in comparison to the background (see Fig. 6).

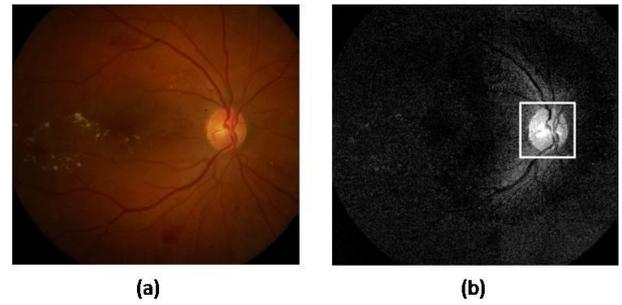

Fig. 6. (a) Original fundus image, (b) its blue channel and sub-image covering the ONH

After generating the sub-images into windows or blocks, the RT is applied to the masked image as noted in Equation (2).

The amplitude of the projection along the diagonal directions ($\theta=45°$, $\theta=135°$, $s=n\sqrt{2}$) is higher than other directions; thus, the peak of RT occurs mostly along diagonal directions. To eliminate the diagonal effect, the sub-images are first masked via a circular mask. The masking process is illustrated in Figs. 7 and 8.

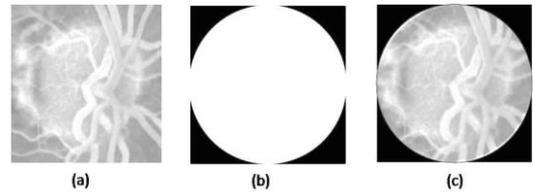

Fig. 7. Masking process in FA: (a) original sub-image, (b) applied mask, (c) masked sub-image

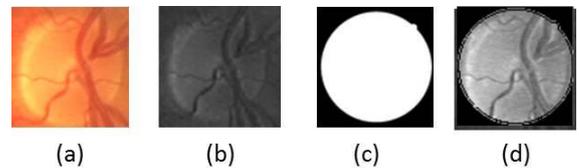

Fig. 8. Masking process in color: (a) original sub-image, (b) blue component, (c) applied mask, (d) masked sub-image

As mentioned earlier, ONH in a sub-image is associated with a prominent peak in the Radon space. Hence, at this stage, it is required to detect peaks (see Fig. 8). Peaks are detected in the Radon space and the profiles in which they occur serve as candidates for ONH. These profiles are further analyzed for validation of candidate ONHs. All sub-images which have a peak profile higher than a predefined threshold are compared. In the results reported in this paper, the threshold was considered to be 0.9 of the largest Radon projection in the sub-image. In other words, initially all Radon projections for one sub-image are found and then the maximum is found. If the peak of RT along all the projection angles is greater than 0.9 of the largest RT projections in the sub-image, that sub-image is regarded as a candidate that may contain the ONH.

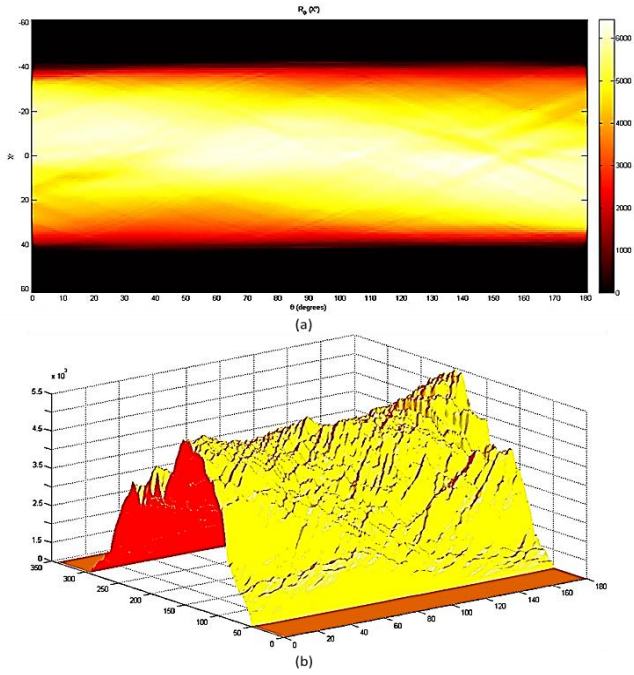

Fig. 9. (a) 2D, and (b) 3D image resulting from Radon transformation of a masked sub-image containing ONH

*Step 3) ONH validation*

The concept of validation is similar to peak measuring along all the projection angles in Radon transformation of a sub-image with a central point pattern (see Fig. 9). Based on this approach, all the sub-images which have a peak profile higher than a predefined threshold get evaluated.

To detect ONH, the RT property for round objects is used. For a round object, RT provides the same profile along all the directions. Due to the roundness of ONH, the profiles related to the projections do not differ much. As a result, one can detect the sub-image that contains the ONH. This is achieved by computing the mean square error (MSE) between the projections. The MSE between the projections inside the blocks is used as a similarity measure for the Radon peaks. In other words, the sub-image which minimizes the MSE among all of its different projections is found (see Fig. 10 and Fig. 11).

The MSE is computed as follows:

$$\text{MSE}(r) = \frac{1}{m \cdot n} \sum_{i=1}^{n} \sum_{j=1}^{m} (R(i,j) - R(r,j))^2 \qquad (4)$$

where $R(i, j)$ denotes the $(i, j)^{th}$ component of the Radon matrix $(R)$, $M$ and $N$ are the dimensions of $R$ and $r$ is the index of the reference column in $R$.

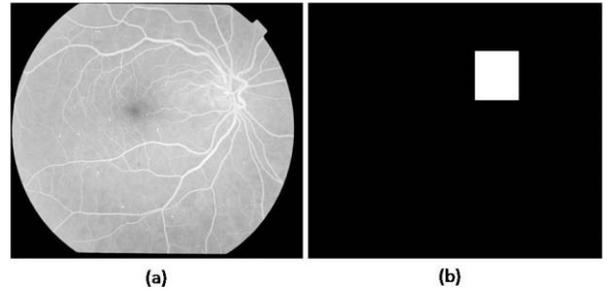

Fig. 10. (a) Original fluorescein angiography fundus image from MUMS-DB, (b) its validated ONH sub-image

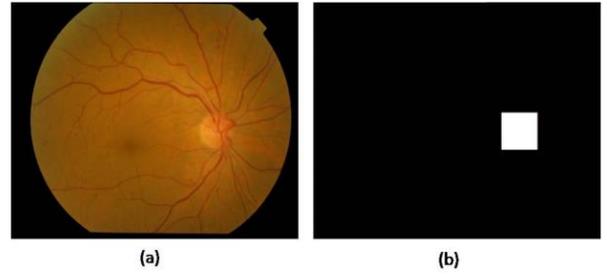

Fig. 11. (a) Original fundus image from MUMS-DB, (b) its validated ONH sub-image

*Step 4) ONH detection*

ONH detection and its changes are quite important in identifying some pathologic conditions. Contrast variance of ONH could generate some artifacts in image segmentation of fundus images. Thus, it is necessary to find ONH and mask it. After the above validation of ONH, the center of the sub-image which touches the ONH is considered to be the center of the ONH. Some sample results in FA and color fundus images from the three databases examined are shown in Figs. 12 through 15.

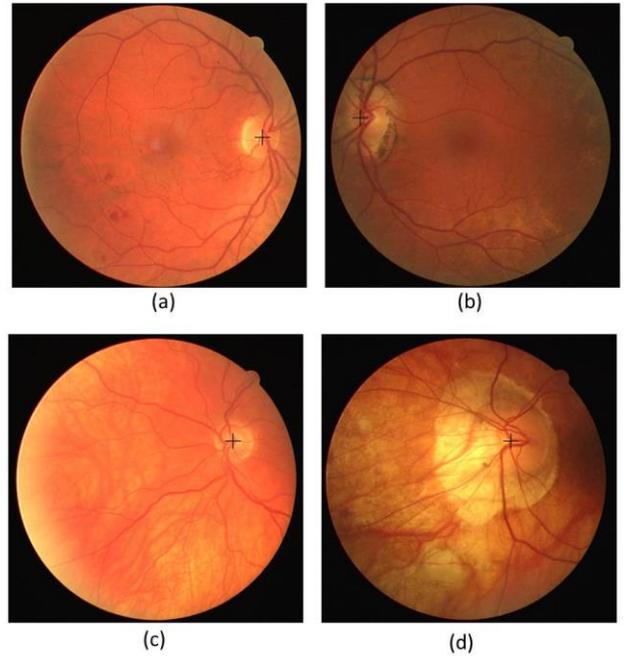

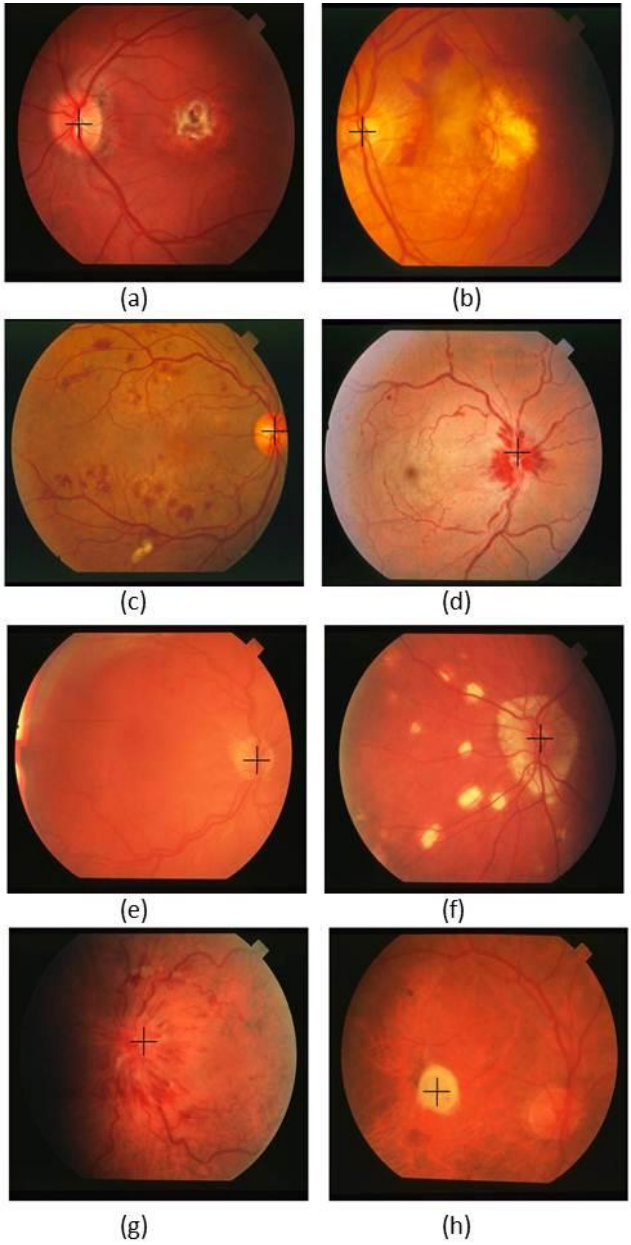

Fig. 12. Outcome of the introduced method applied to sample images of DRIVE database image (black cross represents the estimated ONH center).

Fig. 13. Outcome of the introduced method applied to sample images of STARE database (black cross represents the estimated ONH center): (a)–(g) correct detection, (b) h failure case of the detection

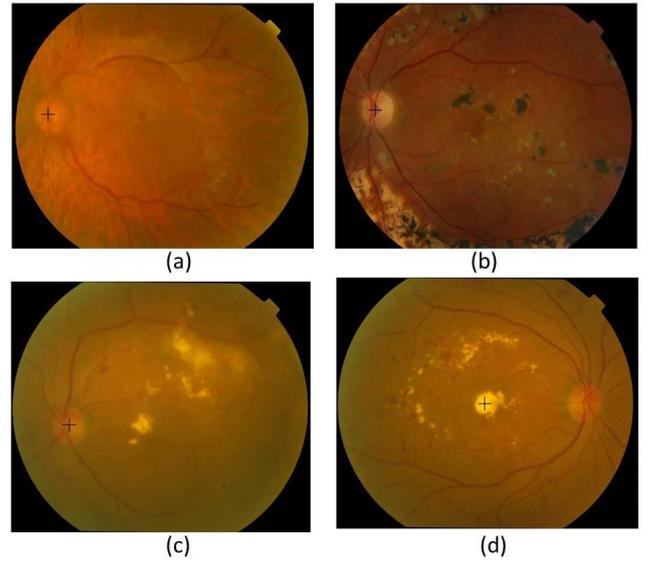

Fig. 14. Outcome of the introduced method applied to sample images of MUMS-DB color database (black cross represents the estimated ONH center): (a)–(c) correct detection, (b) d failure case of the detection

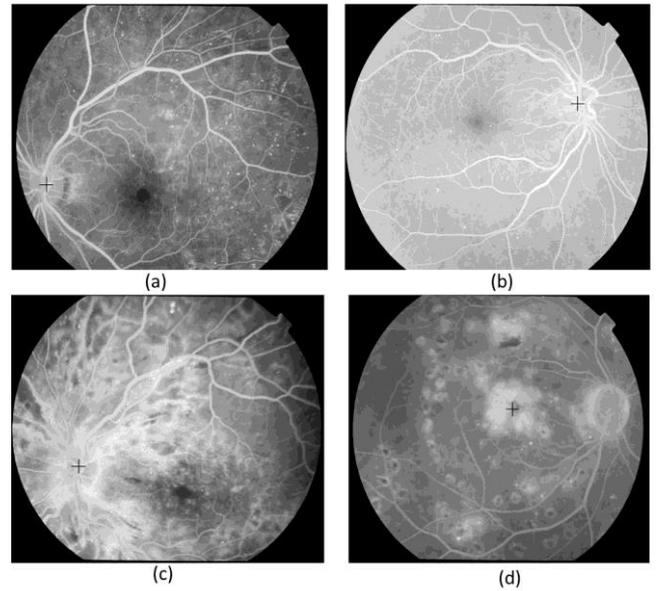

Fig. 15. Outcome of the introduced method applied to sample images of MUMS-DB FA database (black cross represents the estimated ONH center): (a)–(c) correct detection, (b) d failure case of the detection

## C. Computational complexity

The main attribute of our algorithm is its computational efficiency compared with other algorithms. If the input image is size of $M \times N$, the window of size $n$ with the step $s$, the number of operations is as follows:

Fundus area detection: $NM$

Radon Transform: $9MNs^2a$, where $a$ indicates the number of projections in Radon Transform

Validation: $3anc$, where $c$ indicates the number of candidates

Hence, the total number of operations equals *NM* $(1+9as^2)$ + *3anc*.

## IV. RESULTS AND DISCUSSION

Our algorithm was written in MATLAB and applied to the color retinal images in the public DRIVE and STARE databases and to the local database MUMS-DB which include normal and diabetic FA and color retinal images. The DRIVE database consists of 40 images in which 33 images do not have any sign of DR and 7 images show signs of early or mild DR. The image resolution is 564 by 584 pixels acquired using a Canon CR5 3CCD camera with a 45° field of view (FOV). The STARE database consists of 81 fundus images that were used initially by Hoover and Goldbaum [36] to assess their automatic ONH localization technique. The images were captured using a TOPCON TRV 50 retina camera at 35° FOV with an image resolution of 700×605 pixels. This dataset contains 31 images of normal retina and 50 images of diseased retina. The MUMS-DB database consists of more than 2000 retinal images (FA and color). All images were obtained using a TOPCON (TRC-50EX) retinal camera at 50° FOV. Images were mostly obtained from the posterior pole view including optic disc and macula. Their image resolution is 2896 × 1944 pixels.

We examined 120 images for each image type (120 for FA and 120 for color) including 100 cases in different stages (4 stages) of DR, and 20 without DR with no systemic disease or ocular micro-vascular involvement. The selection and diagnosis was done by an ophthalmologist in a blind fashion. The ophthalmologist was asked to mark the ONHs in these images after adding the transparent layer to the images. The ONH locations were saved in similar size files to form the ground truth files. The ground truth files of selected fundus images were also collected. Based on the manual or visual ONH detection, our algorithm found the true locations of ONH in 117 out of 120 color images (97.5%) and 110 out of 120 in FA images (91.3%) of the MUMS-DB database. Some examples of the ONH detection for both the FA and color images of the MUMS-DB database are shown in Figs. 14 and 15.

Additionally, the ONH location was detected correctly in all of the 40 images of the DRIVE database (100%) using our method (see Fig. 12). Our method detected the ONH locations correctly in 78 out of the 81 images of the STARE dataset (96.3%), (see Fig. 13).

TABLE I

DATABASES AND ACCURACIES

| Database | Images | *n* | Accuracy |
|---|---|---|---|
| MUMS-DB (color) | 120 | 313 | 97.5% |
| MUMS-DB (FA) | 120 | 313 | 91.3% |
| DRIVE | 40 | 79 | 100% |
| STARE | 81 | 130 | 96.3% |

The estimated ONH center was considered to be correct if it was positioned within 60 pixels of the manually identified center, as proposed in [10]. The average distance (for the 69 successful images) between the estimated ONH center and the manually identified center was 10 pixels and for the 9 successful images, this distance was 21 pixels in the STARE dataset. In the DRIVE database, the average distance of 40 successful images between the estimated ONH center and the manual segmentation was 13 pixels. Finally, the average distance of the successful images in the MUMS-DB database was 14 pixels. The cases in which the ONHs were not correctly detected were due to uneven round-shaped illumination that biased the ONH candidates (see Figs. 13(h), 14(d), and 15(d)).

Table II compares the accuracies of the introduced method and two representative existing methods which have been reported to have the highest detection rates and at the same the fastest detection times.

TABLE II

COMPUTATIONAL COMPLEXITY COMPARISON OF ONH DETECTION METHODS AND DETECTION RATES FOR STARE DATABASE

| Methods | Detection Rate | No. of failed images | Accuracy (pixels) | Computational complexity in terms of approximate number of operations | Notations |
|---|---|---|---|---|---|
| Foracchia et al.[10] | 97.5% | 2 | 23 | $24MNW^2+180IV$ | $W = 16$<br>$V = 300$<br>*I*: no of iterations in simulated annealing |
| Youssif et al. [11] | 98.8% | 1 | 26 | $MN(W_1^2+2W_2^2+24W_3^2+15)+324VW_4^2$ | $W_1 = 40$<br>$W_2 = 80$<br>$W_3 = 15$<br>$W_4 = 40$<br>*V*: Number of vessel pixels |
| Mahfouz [33] | 92.6% | ?? | 14 | $MN(10+V)+2MW^2+4W^2$ | *W*: ONH size (130)<br>*V*: Vessel width |
| Introduced method | 96.3% | 3 | 10 | $NM(1+9as^2) + 3anc$ | $s = 4$<br>$a = 12$<br>$n = 130$ |

$M$ and $N$ are 605 and 700, respectively, in the STARE dataset. As can be seen from Table II, the number of operations in our algorithm is much less than Foracchia et al. [10] and Youssif et al. [11] which have reported slightly higher detection rates, while the number of operations is higher than Mahfouz [33] which has reported lower detection rates. Basically, our algorithm runs the fastest among those algorithms that provide relatively high detection rates.

In terms of actual processing time, our algorithm processes an image in the STARE database in only 4.1 sec on a 2MHz Intel Core i5. The method of Youssif *et al.* [11] as reported in [33] takes 3.5min on a 2MHz Intel Centrino 1.7 CPU. The method of Foracchia *et al.* [10] as reported in [33] takes 5 min for vessel extraction on a Sun SPARCstation 20 and 2 min for detection on a 2MHz Intel Pentium IV. The only faster algorithm reported in the literature is the one by Mahfouz [33] which takes just 0.46 sec on a 2.66 Intel Core2Due, however, as noted above, its reported detection rate is lower than our method with ONH center error of 14 pixels.

Furthermore, processing of high resolution images in the MUMS-DB (2896 × 1944 pixels) took only around one minute. Our detection algorithm was also found to be scalable in terms of image resolution. We tested it on half-sized retinal images of the three databases. It was found that the ONH detection accuracy still reached 100% for DRIVE, 96.3% for STARE, and 95.9% for MUMS-DB color databases while the processing time was speeded up significantly, that is more than 12 times faster than when using the original resolution.

Note that our method unlike some studies, e.g. [22, 28, 29], does not need to segment retinal vessels for the detection of ONH, rather it relies on the size, intensity, and roundness information. As a result, all the failed retinal images reported in [11] and [10] (i.e., (b) and (d) images in Fig. 13) were correctly detected by our method.

Finally, it is worth stating that our method still has difficulty with a very small or uncommon number of retinal images whose ONH is darker than the surrounding pixels since it is designed based on the assumption that ONH is more or less brighter than the surrounding retinal pixels.

## V. CONCLUSION

In this paper, we have discussed an automated algorithm to detect Optical Nerve Head in a computationally efficient manner. Our algorithm uses a unique combination of Radon Transform and multi-overlapping windows to achieve a robust approach to various conditions such as Diabetic Retinopathy in color fundus images as well as in fluorescein angiography images. Three databases were examined and it was found that our algorithm exhibited high detection rates while achieving higher processing rates than those existing algorithms with comparable detection rates.

ACKNOWLEDGEMENT

This work was partially supported by the graduate student grant no. 87394 from Mashhad University of Medical Sciences (MUMS). The authors would like to thank the staffs of Khatam-Al-Anbiya Eye Hospital for their assistance with the image acquisition.